\documentclass[12pt]{article}

\newcommand{\br}{\hskip .25cm/\hskip -.25cm}
\begin{document}

\sloppy
\begin{flushright}{SIT-HEP/TM-9}
\end{flushright}
\vskip 1.5 truecm
\centerline{\large{\bf Baryon number violation, baryogenesis}}
\centerline{\large{\bf and defects with extra dimensions}}
\vskip .75 truecm
\centerline{\bf Tomohiro Matsuda
\footnote{matsuda@sit.ac.jp}}
\vskip .4 truecm
\centerline {\it Laboratory of Physics, Saitama Institute of
 Technology,}
\centerline {\it Fusaiji, Okabe-machi, Saitama 369-0293, 
Japan}
\vskip 1. truecm
\makeatletter
\@addtoreset{equation}{section}
\def\theequation{\thesection.\arabic{equation}}
\makeatother
\vskip 1. truecm

\begin{abstract}
\hspace*{\parindent}
In generic models for grand unified theories(GUT), various types of
baryon number violating processes are expected when quarks and leptons
propagate in the background of GUT strings.
On the other hand, in models with large extra dimensions, the baryon number
violation in the background of a string is not trivial because it
must depend on the mechanism of the proton stabilization.
In this paper we argue that cosmic strings in models with extra
 dimensions can enhance the baryon
 number violation to a phenomenologically interesting level, if the proton
decay is suppressed by the mechanism of localized wavefunctions.
We also make some comments on baryogenesis mediated by 
cosmological defects. 
We show at least two scenarios will be successful in this direction.
One is the scenario of leptogenesis where the required lepton number
conversion is mediated by cosmic strings, and the other is the
 baryogenesis from the decaying cosmological domain wall.
Both scenarios are new and have not been discussed in the past.
\end{abstract}

\newpage
\section{Introduction}
\hspace*{\parindent}
Although the quantum field theory made a great success, there is no
consistent scenario in which the quantum gravity is included.
The most promising framework that could help in this direction would be
the string theory, whose consistency is maintained by the requirement of
additional dimensions.
At first the sizes of extra dimensions had been assumed to be $M_p^{-1}$,
however it has been observed later that there is no obligation
to consider such a tiny compactification radius\cite{extra_1}.
The large compactification scale may solve or weaken the 
traditional hierarchy problem.
If we denote the volume of the $n$-dimensional compact space by $V_n$,
the observed Planck mass $M_p$ is obtained by the relation
$M_p^2=M_{*}^{n+2}V_n$, where $M_{*}$ is the fundamental scale of gravity.
If one assumes more than two extra dimensions, $M_*$ can be very
close to the electroweak scale without conflicting any observable
bounds.

Although such a low fundamental scale considerably improves the standard
situation of the hierarchy problem, the scenario requires some degrees of
fine tuning, which is apparently a modification of old hierarchy problems.
The most obvious example of such fine tuning is the largeness of the
quantity $V_n$.
There are other aspects of fine tuning, which are 
common to conventional scenarios of grand unified theories.
In particular, the problem of obtaining sufficient baryon number
asymmetry of the Universe while keeping protons stable
should be one of the most serious problems in models with low
fundamental scale. 
In standard GUT theories, the proton stability is usually achieved by
increasing the mass of the GUT particles that mediate the baryon
number violating interactions.
At the same time, the GUT-scale heavy particles assist the
non-equilibrium production of the baryon number asymmetry. 
However, in theories with low fundamental scale, the suppression can not
be achieved by merely increasing the mass scale.
In this respect, some non-trivial mechanism is needed to solve this problem.

Recently an interesting mechanism was suggested in
ref.\cite{proton}, where 
a dynamical mechanism of localizing fermions on the thick wall is
adopted to solve the problem of fast proton decay.
In this scenario, leptons and baryons are localized at displaced
positions in the extra space, where the smallness of their interaction is 
insured by the smallness of the overlap of their wavefunctions along the
extra dimension.

On the other hand, to explain the observed baryon asymmetry of the
Universe, baryon number violating interactions must have been
effective but non-equilibrium in the early Universe, because
the production of net baryon asymmetry requires baryon number
violating interactions, C and CP violation and a departure from the
thermal equilibrium\cite{sakharov}.
If the fundamental mass scale is sufficiently high, 
the first two of these ingredients are naturally contained in
conventional GUTs.
The third can be realized in an expanding universe where it is not
uncommon that interactions come in and out of equilibrium, producing the
stable heavy particles or cosmological defects.
In the original and simplest model of baryogenesis\cite{original, decayb},
a heavy GUT gauge or Higgs boson decays out of equilibrium producing
a net baryon asymmetry.

In models with large extra dimensions, however, the situations are rather
involved because of the low fundamental scale and the requirement for the
low reheating temperature. 
Such a low reheating temperature makes it much more difficult to produce
the baryon asymmetry while achieving the proton 
stability in the present Universe.\footnote{Aspects of baryogenesis 
with large extra dimensions are already discussed by many authors.
For example, in ref.\cite{mazumdar, matsuda_thermalbrane}, it is argued
that the Affleck-Dine mechanism can generate adequate baryogenesis.
In ref.\cite{dvali}, the global-charge non-conservation due to quantum
fluctuations of the brane surface is discussed. 
The baryogenesis by the decay of heavy X particle is discussed in
ref.\cite{thermalB, matsuda_defect}. 
In ref.\cite{four-point}, it is argued that a dimension-6 proton decay
operator, suppressed today by the mechanism of quark-lepton separation
in extra dimensions can generate baryon number if one assumes that this
operator was unsuppressed in the early Universe due to a time-dependent
quark-lepton separation.}  
In this respect, it is very important to propose ideas to
enhance the baryon number violating interactions that can take place
even in the models with low reheating temperature.

In this paper we propose a mechanism where the enhancement of the baryon
number violating interaction is realized by the cosmological defect.

The plan of our paper is the following.
In section 2 we show how to enhance the baryon number violating
interaction in the background of a string.
Although the mechanism might seem similar to the one in the standard GUT
string, there is a crucial difference.
In the standard scenario of the GUT string, the cross section for the
baryon number violating interactions mediated by the string is usually
enhanced by the factor $\left(\frac{M_{GUT}}{M_{proton}}\right)^2$. 
We stress that one cannot reproduce this enhancement in models with
extra dimensions merely extrapolating the analyses on the standard GUT
string. 
We also make a brief comment on the scenario for leptogenesis with low
reheating temperature.
If the maximum temperature is lower than the temperature for the
electroweak phase transition, sphalerons cannot convert
existing leptons 
into baryons, which is a serious problem for leptogenesis.
Even if sphalerons are not activated, strings with enhanced baryon
number violation can mediate the baryon number production.
In section 3 we comment on baryogenesis from the decaying
cosmological defect.
Cosmic strings are not effective in generic cases, but domain walls are
promising.

\section{Defects and enhanced baryon number violation}
\hspace*{\parindent}
Our first task is to review the old issues of enhanced baryon
number violation due to conventional GUT strings.
Then we extend these analyses to models with extra dimensions, where the
fast proton decay is suppressed in the present Universe due to the
localized wavefunctions along extra dimension.
One can easily find why the naive application
cannot reproduce the enhancement of the baryon number violation that
appears in standard GUT strings.
We solve this problem, and construct strings with enhanced baryon number
violation in models with extra dimensions.
Our idea is based on the idea of ref.\cite{matsuda_defect}. 
We stress that our mechanism works without GUT symmetry, but
some extensions are required for the standard model.

\subsection{Enhanced baryon number violation in standard GUT strings}
\hspace*{\parindent}
In the interior of a string formed after GUT symmetry breaking, there
are fields that carry both baryon and lepton number.
Thus a quark comes into the core of such strings will interact with the
background core fields, scattered to emerge as a lepton, and vice versa.
A GUT string therefore is a candidate source for baryon number violating 
processes in the early Universe.
This reminiscents of the Rubakov-Callan effect\cite{monopole}, where the
baryon number violating interaction is mediated by monopoles.

To be more specific, inside the core of the string there are quark-lepton
transition mediated by $X^{\mu}$ and $Y^{\mu}$ gauge bosons, or
Yukawa couplings to charged scalar fields $\phi_X$ that may condensate
within the string core\cite{superconducting}.

Denoting the scalar field that forms the string by $\phi_{string}$,
a perturbative calculation of the scattering cross-section per unit
length $(d \sigma/ dl)$ for a scalar particle $\phi$ coupled as 
$\frac{1}{2} \lambda' |\phi_{string}|^2 \phi^2$ reveals that \cite{BDM}
\begin{equation}
\left(\frac{d \sigma}{dl}\right)
\sim \left(\frac{\lambda'}{\lambda}\right)^2 E^{-1},
\end{equation}
where $\lambda$ denotes the self-coupling constant of $\phi_{string}$.

Fermions have different couplings and different phase spaces, and
require more discussions.
With a simple coupling, $g \phi_{string}^{*}\overline{\psi}^{c} \psi$,
the cross-section per unit length is calculated as \cite{BDM},
\begin{equation}
\left(\frac{d \sigma}{dl}\right)
\sim \left(\frac{g}{\lambda}\right)^2 \frac{E}{M_{GUT}^2},
\end{equation}
which is ``not enhanced''.
However, there is a crucial factor in the calculation for gauge strings.
If the charge of the string field $\phi_{string}$ is $e$, the resulting
cross-section depends strongly on the ratio $q/e$, where $q$ is
the charge of the scattered particle under the generator of the broken 
$U(1)$.
The elastic scattering cross-section depends crucially on the ratio
$q/e$, since it controls the amplification of the scattering
wave-function at the core of the string\cite{string_AB}.
In general, the cross section also depends on the type of the
interaction, whether it is mediated by a scalar or a vector field.
A vector field produces its maximum effect for integer $q/e$, then the
baryon-number violating scattering cross section per unit length
$\sigma_{\br{B}}$ becomes $\sim E^{-1}$.
The baryon-number violating cross-section from a core scalar field takes
it maximum value for half-integer $q/e$, then it reaches $\sim E^{-1}$.

We shall consider a simple example.
Here we follow the settings and analysis in ref.\cite{AMW} and show the
outline of the argument.
We consider a straight string formed by a $U(1)_S$-charged scalar field
$\phi_{string}$, which condensates outside the string,
i.e., $<\phi_{string}>\simeq \eta$.
This $U(1)_S$ is orthogonal to electromagnetic $U(1)_{em}$.
We include a second scalar field $\phi_X$ that is not charged under
$U(1)_S$, carries baryon number, and condensates within the string core.
The Lagrangian includes Yukawa interactions
\begin{equation}
\label{lag1}
{\cal L}= -\lambda(\phi_X \overline{\chi}\psi+\phi_X^*\overline{\psi}\chi)
\end{equation}
where $\psi$ and $\chi$ are both fermions of $U(1)_S$ charge.
In ref.\cite{AMW}, the cross section for the production of $\chi$
in the scattering of $\psi$ fermions off a string is calculated.
The Dirac equations for $\psi$ and $\chi$ have off-diagonal element
$\lambda<\phi_X>$, which vanishes outside the string but becomes non-zero
in the string core.
Schematically, the Direc equations are written by
\begin{equation}
\left(
\begin{array}{cc}
i \br{\partial} - i\br{A}-m_{\psi} & \lambda<\phi_X>^{*}\\
\lambda<\phi_X>^{*} & i \br{\partial} - i\br{A}-m_{\chi}
\end{array}
\right)
\left(
\begin{array}{c}
\psi\\
\chi
\end{array}
\right)
=0
\end{equation}
where 
\begin{equation}
A=\frac{1}{g r^2}\left(
\begin{array}{c}
0\\
-y\\
x
\end{array}
\right)
,
\,\,\,
\lambda<\phi_X>=
\left\{
\begin{array}{cc}
0, & for \,\,[r>R]\\
v, & for \,\,[r<R].
\end{array}
\right.
\end{equation}
The result is interesting, since the cross sections for
these processes are generically enhanced by the large factor over the naive
geometrical cross section.
They have explicitly calculated the cross section for quarks and
leptons, which are denoted by $\psi$ and $\chi$ in eq.(\ref{lag1}),
and found that the enhance factor becomes up to 
$\left(\frac{M_{GUT}}{M_{proton}}\right)^2$.
In ref.\cite{AMW}, the baryon-number violating
processes including couplings to core gauge fields $X^{\mu}$ and
$Y^{\mu}$ are also calculated and found that these scattering cross
sections are enhanced by the same factor.

With this in mind, it is natural to ask whether a similar effect occurs
for cosmic strings in models with extra dimensions.
In the models with localized wavefunctions along the extra dimension, the
baryon-number breaking interactions, such as 
$\lambda' \phi_X\overline{\psi}\chi$ must be suppressed exponentially 
in the true vacuum so as to ensure the proton stability.
In this case, even if the charged scalar $\phi_X$ condenses in the
core of the string, $\lambda'<\phi_X>$ remains negligibly small.
Thus in the naive setup, the baryon number violating interaction
mediated by the string with extra dimensions is exponentially
suppressed and phenomenologically negligible.

In the next subsection we show that we can improve this disappointing
result if the idea of ref.\cite{matsuda_defect} is taken into account.

\subsection{Defects in models with extra diemnsions}
\hspace*{\parindent}
Now we want to construct viable models for cosmic strings
that can mediate baryon number violation in models with extra dimensions.
Although the TeV scale unification is already discussed by many
authours\cite{TeV_GUT_power,TeV_GUT_4D}, a naive extrapolation of the
standard GUT string cannot madiate baryon-number violating interactions,
from the reason that we have discussed above.
In the following, we do not assume explicit
realization of GUT, since one may include baryon-number breaking
couplings without assuming GUT embeddings.
These baryon-number violating couplings must be suppressed in
the vacuum to prevent fast proton decay. 

In ref.\cite{matsuda_defect}, we have proposed a new scenario of baryon
number violation, which can be utilized to solve this dilemma.
To show the elements of our idea, here we limit ourselves to
constructions with fermions 
localized within only one extra dimension\cite{proton} and show how the
tiny couplings can be enhanced in the defect.

To localize fields in the extra dimension, it is necessary to 
break higher dimensional translation invariance, which is
accomplished by a spatially varying expectation value of the
five-dimensional scalar field 
$\phi_{A}$ of the thick wall along the extra dimension.
If the scalar field $\phi_{A}$ couples to the five-dimensional 
fermionic field $\psi$ through the five-dimensional Yukawa interaction
$g \phi_A \overline{\psi}\psi$, whose expectation value $<\phi_A>$ varies 
along the extra dimension but is constant in the four-dimensional world,
it is possible to show that the fermionic field localizes at the place
where the total mass in the five-dimensional theory vanishes.
For definiteness, we consider the Lagrangian
\begin{eqnarray}
{\cal L} &=&\overline{\psi_{i}}\left(i \br{\partial_5} +g_{i}\phi_{A}(y) 
+m_{5,i}
\right)\psi_{i}\nonumber\\
&&+\frac{1}{2}\partial_{\nu}\phi_{A} \partial^{\nu}\phi_{A} \nonumber\\
&&-V(\phi_{A}),
\end{eqnarray}
where $y$ is the fifth coordinate of the extra dimension.
For the special choice $\phi_{A}(y)=2\mu^2 y$, which corresponds to 
approximating the kink with a straight line interpolating two vacua,
the wave function in the fifth coordinate becomes gaussian centered
around the zeros of $g_{i}\phi_{A}(y)+m_{5,i}$.
It is also shown in ref.\cite{extra_1} that a left handed chiral
fermionic field in the four-dimensional representation can result from
the localization mechanism.
The right handed part remains instead delocalized in the fifth
dimension.
When leptons and baryons have the five-dimensional masses
$m_{5,l}$ and $m_{5,q}$, the corresponding localizations are
at $y_l=-\frac{m_{5,l}}{2g_l \mu^2}$ and $y_q=-\frac{m_{5,q}}{2g_q
\mu^2}$, respectively.
The shapes of the fermion wave functions along the fifth direction are 
\begin{eqnarray}
\Psi_l (y)&=&\frac{\mu^{1/2}}{(\pi/2)^{1/4}}
exp\left[-\mu^2 (y-y_l)^2\right]\nonumber\\
\Psi_q (y)&=&\frac{\mu^{1/2}}{(\pi/2)^{1/4}}
exp\left[-\mu^2 (y-y_q)^2\right].
\end{eqnarray}
Even if the five-dimensional theory violates both baryon and lepton
number maximally, the dangerous operator in the effective
four-dimensional theory is safely suppressed.
For example, we can expect the following dangerous operator in the
five-dimensional theory,
\begin{equation}
{\cal O}_5\sim \int d^5 x \frac{QQQL}{M_*^3}
\end{equation}
where $Q,L$ are five-dimensional representations of the fermionic fields.
The corresponding four-dimensional proton decay operator is obtained by
simply replacing the five-dimensional fields by the zero-mode fields and
calculating the wave function overlap along the fifth dimension $y$.
The result is
\begin{equation}
{\cal O}_4 \sim \epsilon \times \int d^4 x \frac{qqql}{M_{*}^2},
\end{equation}
where $q,l$ denotes the four-dimensional representation of the chiral
fermionic field.
The overlap of the fermionic wavefunction along the fifth dimension is
included in $\epsilon$,
\begin{equation}
\epsilon \sim \int dy \left(e^{-\mu^2 (y-y_q)^2}\right)^3 e^{-\mu^2 (y-y_l)^2}.
\end{equation}
For a separation $r=|y_q - y_l|$ of $\mu r =10$, one can obtain
$\epsilon \sim 10^{-33}$ which makes this operator safe even for 
$M_*\sim$TeV.
On the other hand, however, this suppression prevents the required
baryon-number violating interaction of the form
\begin{eqnarray}
\label{Yukawa_X}
{\cal L}_{\phi_X lq}&=& \lambda'\phi_X lq,
\end{eqnarray}
where $\lambda'$ is exponentially suppressed.

To construct defect configurations, we extend the above idea to include
another scalar field $\phi_B$ that determines the five-dimensional mass
$m_5$. 
This additional field $\phi_B$ determines
the position of the center of the fermionic wavefunction along the fifth
dimension. 
We assume that $\phi_B$ does {\it not} make a kink
configuration along the fifth dimension, but {\it does} make a defect
configuration in the four-dimensional spacetime. 
For definiteness, we consider the Lagrangian
\begin{eqnarray}
{\cal L} &=&\overline{\psi_{i}}\left(i \br{\partial_5} +g_{i}\phi_{A}(y)
+m(\phi_{B})_{5,i}
\right)\psi_{i}\nonumber\\
&&+\frac{1}{2}\partial_{\nu}\phi_{k}\partial^{\nu}\phi_{k}\nonumber\\
&&-V(\phi_{k}),
\end{eqnarray}
where indices represent $i=q,l$ and $k=A,B$.
Here $\phi_{A}$ makes the kink configuration along the fifth dimension
while $\phi_B$ develops defect configuration in the four-dimensional
spacetime.
We note that $\phi_B$ plays the role of $\phi_{string}$ in the
previous arguments. 

Now we consider the simplest case where $m_{5,i}$ are given by
$m(\phi_{B})_{5,i}=k_{i} \phi_{B}$, and the potential for $\phi_B$
is given by the double-well potential of the form; 
\begin{equation}
V_{B}=-m_B \phi_{B}^2 +\lambda_B \phi_B^4.
\end{equation}
In this simplest example, because of the effective $Z_2$ symmetry,
the resultant defect is the cosmological domain wall.
One can easily extend the model to have the string or the monopole
configuration in four-dimensional spacetime, if the appropriate symmetry
is imposed on the scalar field $\phi_B$.
For example, one can consider the form
\begin{equation}
m(\phi_B)_{5,i}=k_{i}\frac{|\phi_{B}|^2}{M_*}
\end{equation}
where $\phi_B$ is charged with $U(1)_S$.

In any case, the center of the fermionic wavefunction in the fifth
dimension can be shifted by the defect configuration in the
four-dimensional spacetime.
Because of the volume factor suppression, the largest contribution is
expected in 
the quasi-degenerated vacuum surrounded by the cosmological domain-wall
that interpolates between $\phi_B = \pm v$.
Let us consider the case where the wavefunctions of quarks and
leptons are localized at the opposite side of the $\phi_A$ kink.
It can be realized if their couplings to $\phi_B$ have opposite signs.
In the core of the $\phi_B$ string, the centers of the
quarks and leptons move toward the origin.
Then the distances between quarks and leptons become $r=0$ in the
core of the string, which drastically modifies the magnitude of the
baryon number violating interactions mediated by the string.

If one assumes GUT unification, the origin of such strings becomes
clear.
For the simplest example, we shall follow ref.\cite{proton, TeV_GUT_4D} 
and show how one can realize the localization in GUT embeddings.
The embedding of SM gauge groups in some grand unified group is an
interesting issue.
However, the most serious obstacle for the low energy unification
is the problem of proton stability, which prevents
such embedding at low GUT scale.
If the quarks and leptons are unified into the $SU(5)$ multiplets
$\overline{5}$ and $10$, then the heavy gauge bosons $X$ and $Y$ must
mediate the fast proton decay.
In ref.\cite{proton, TeV_GUT_4D}, GUT models with stable proton are
constructed by using the mechanism of localized wavefunctions.
In the bulk, Dirac masses residing in quintuplets and
decuplets can be split as a result of GUT symmetry breaking  according
to the following equations:
\begin{eqnarray}
\overline{5}\left(<\Sigma>+M_5 + \phi_A\right)5&=&0\nonumber\\
\overline{10}\left(<\Sigma>+M_{10} + \phi_A\right)10&=&0.\nonumber\\
\end{eqnarray}
where $<\Sigma>=v\times diag(2,2,2,-3,-3)$.
In this model any transition between quarks and leptons in four
dimensions is naturally suppressed by the exponential factor.
In this GUT example, $<\Sigma>$ plays the role of $m_{5,i}$
in the previous discussions. 
Although it requires some extensions to larger gauge groups than $SU(5)$,
GUT string can be formed as in the standard GUT scenarios, and
the GUT symmetry is restored inside the core of the string.
Thus in the GUT models for wavefunction localization, it is natural to
expect that the positions of the fermionic wavefunctions along the fifth
dimension are modified in the background of the GUT defects in the
four-dimensional spacetime.  
 
Although it seems rather difficult to produce these defects merely by the
thermal effect after inflation, nonthermal effect may create such
defects during preheating period of inflation.
Nonthermal creation of matter and defects has raised a remarkable
interest in the last years.
In particular, efficient production of such products during the period
of coherent oscillations of the inflaton has been studied by many
authors\cite{PR}.\footnote{
There is an another possibility that the defects are generated after the
first brane inflation, while the reheat temperature after the second thermal
brane inflation is kept much lower than the electroweak
scale\cite{thermalbrane}. }

Thermal effect becomes important if one considers the
supersymmetric extension with intermediate mass scale, i.e., $M_* >>TeV$, 
where the position of
the localized matter field can be parameterized by the flat direction.
In this case, one can expect thermal symmetry restoration at the
temperature much lower than the cut-off scale, which is accessible in
realistic scenarios.
During thermal symmetry restoration, if the five-dimensional mass
terms $m_{5,i}$ is determined by a field $\phi_B$ that parameterizes the flat
direction, the center of the wavefunction is shifted along extra
dimension, which results in the huge enhancement in the exponential
factor. 
Then the baryon number violation can become effective till
$T\sim O(10^2 GeV)$.
In this case, the defect formation is not the sole candidate that
enhances the baryon number violation.

\section{Comments on baryogenesis}
\hspace*{\parindent}
In this section we make some comments on baryogenesis that is mediated
by cosmological defects in models with extra dimensions.

\subsection{Particle scattering and baryon number violation}
\hspace*{\parindent}
The effect of the baryon number violation is greatest when the density of
the string is greatest, which in general occurs soon after the phase
transition. 
Unlike conventional GUT strings, the string with extra dimensions is
formed at much lower energy scale, and remains friction-dominated. 
If the string is the effective source of the baryon number violation,
any existing baryon number asymmetry in the Universe may be washed out.
On the other hand, if there is no effective source of the baryon number
violation other than the Yukawa coupling $\phi_X ql$ that is 
enhanced by the string, and also if the reheating temperature is too low
to activate the sphalerons, the string cannot washout the existing
asymmetry, but converts the existing leptons into baryons, and vice versa.
In this case the strings can be utilized for leptogenesis with low reheat
temperature.
\footnote{
Leptogenesis in theories with extra dimensions is already discussed in
ref.\cite{lepto_extra}.
In this scenario the reheating
temperature should be higher than 10 GeV or so, which gives rise to
constraints on the model parameters and the quantum gravity scale.
The observed baryon asymmetry can still be generated by
out-of-equilibrium, exponentially suppressed B+L-violating sphaleron
interactions.}

Denoting the correlation length of the string network by $\xi(t)$, 
and the baryon-number violating cross section per unit length by
$\sigma_{\br{B}}$, one obtains\cite{stringbaryo}
\begin{equation}
\frac{d n_B}{d t} \simeq -\overline{v}\sigma_{\br{B}}\frac{1}{\xi^2}n_L,
\end{equation}
where $n_B(n_L)$ is the baryon(lepton) number density and $\overline{v}$
is the thermally averaged relative velocity of the particles and
strings, which is of order $1$.
We can take
$\sigma_{\br{B}}\sim T^{-1}$ and $\xi(t)\sim t^p$, where $p=5/4$
in the friction-dominated phase\cite{stringbaryo}.

In models with large extra dimensions, the lepton
number will be produced by the decay of the relics of the inflation, 
such as the Affleck-Dine field\cite{Affleck-Dine, matsuda_thermalbrane,
mazumdar},
or the inflaton\cite{thermalB}, which will have large densities.
In this case one can naturally assume that the lepton number production
and the formation of the string occur at almost the same time,
which makes our scenario plausible.
If the lepton number production occurs before the string formation,
the conversion is promising because the scattering is greatest just
after the string formation.
On the other hand, if the strings are formed before the lepton number
production, the conversion is not trivial.
The conversion is greatest just after the time of the lepton production,
$t_L$.
Then one can obtain the ratio
\begin{equation}
\frac{\Delta n_B}{n_L}|_{t_L} \simeq \frac{1}{T(t_L)} 
\left(\frac{1}{\xi_0\left(\frac{t_L}{t_0}\right)^p}\right)^2 t_L,
\end{equation}
where $t_0$ and $\xi_0$ are the time of the string formation and the
initial correlation length of the string, respectively.
The scattering is important when
$\frac{\Delta n_B}{n_L}|_{t_L} \sim O(1)$.
Using the above relations, one can easily find that the scattering
process is activated just after the string
formation and then lasts for a short period.
If the sphalerons and other baryon-number violating interactions are not
activated at the time of the string formation,  
which may occur in the scenario of defect formation 
during preheating\cite{PR}, these strings can convert 
existing leptons into baryons without washout.
We believe that our idea can help to solve the
difficulty in leptogenesis for models with large extra dimensions
related to the low reheat temperature that suppresses the 
sphaleron-mediated interactions.\footnote{In ref.\cite{nonth_riott},
sphaleron transitions that may proceed during matter-dominated phase 
is discussed.
In our model, the reheating process must be different from the one
discussed in ref.\cite{nonth_riott} to avoid washout of the existing
asymmetry.}  

\subsection{Baryogenesis from collapsing strings}
\hspace*{\parindent}
At first we shall review the basics of the standard
scenario of baryon number generation from the decaying GUT strings.
Baryon number
asymmetry produced by strings are important when the number density of
existing heavy $X$ particle is less than the one produced by the
decaying strings.
In the standard GUT baryogenesis it is well known that the predicted
baryon to entropy ratio is exponentially suppressed if $T_d$, which is
the temperature at the time when the baryon number violating processes
fall out of thermal equilibrium, is less than the mass of the heavy X
particle $m_X$. 
For example, if the temperature $T_d$ at the time $t_d$ is greater than
the mass $m_X$ of the superheavy particles, then it follows that
$n_X\sim s$.
However, if $T_d<m_X$, the number density of $X$ is diluted
exponentially, which results in the exponential suppression of the
produced baryon to entropy ratio:
\begin{equation}
\frac{n_B}{s}\sim \epsilon \frac{1}{g^*}\lambda^2
 exp\left(-\frac{m_X}{T_d}\right),
\end{equation}
where $g^*$ is the number of spin degrees of freedom in thermal
equilibrium at the time of the phase transition, $\lambda$ is the
coupling constant of the baryon number violating process, and
$\epsilon$ demotes the effective CP violation. 
If the exponential suppression is tiny, the standard GUT
baryogenesis mechanism is ineffective.

However, topological defects may solve this
problem\cite{stringbaryo} in standard GUT.
The collapse of cosmic string can produce baryon asymmetry
if the $\phi_X$-boson, which couples to both baryons and leptons through
Yukawa coupling of the form $\lambda \phi_X ql$, forms or couples to the
cosmic string.
The important ingredient in the quantitative calculation is the time
dependence of the correlation length $\xi(t)$, which parameterizes the
separation between defects.
Just after the phase transition, the separation is expected to be 
$\xi(t_0)\sim \lambda^{-1}\eta^{-1}$.
The time period of relevance for baryogenesis, $\xi(t)$ approaches the
Hubble radius according to the following equation\cite{xi_t_dependence}
\begin{equation}
\xi(t)\simeq \xi(t_0)\left(\frac{t}{t_0}\right)^{\frac{5}{4}}.
\end{equation}
After some algebra, one can obtain the baryon to entropy ratio:
\begin{equation}
\frac{n_B}{s}\sim \lambda^2 \frac{T_d}{\eta}\frac{n_B}{s}|_0
\end{equation}
where $\frac{n_B}{s}|_{0}$ is the unsuppressed value of GUT
baryogenesis, and $\eta$ denotes the square root of the energy density
per length. 
Thus in the standard GUT models,
one can see that the superheavy particles produced by the succeeding
decay of the cosmic strings can overcome the exponential
suppression.

The domain wall, which seems much more effective for producing particles,
is ruled out by the cosmological requirement.\footnote{
However, we should stress that in models with extra dimensions the mass
scale of the domain wall is much lowered, thus they can survive till the
time period of relevance for baryogenesis.
We shall discuss on this issue in the next subsection.}

Then what is the difference between the defect mediated baryogenesis in the
standard GUT and the one with extra dimensions?
The most obvious difference is the reheating temperature, which must be
much lower than the scale of the string formation.
In scenarios for string mediated baryogenesis, the greatest contribution
appears just after they have formed.
Regarding to the baryon number production by the decaying strings,
later contribution cannot exceed the initial one, because of the rapid
growth of their correlation length and the dilution by the redshift.
Then the low reheating temperature is problematic, since there should be
a large amount of dillution by the decaying field in the time period 
between string formation and reheating.\footnote{
We shall consider two separated situations.
In one case\cite{nonth_riott}, preheating is
not effective and the maximum
temperature during reheating can be much greater than the reheating
temperature. 
This may happen when the reheating is far from being an instantaneous
process, and the decay products of the relevant scalar field thermalize
rapidly before the Universe becomes radiation dominated.
They have calculated the dillution factor of the baryon to entropy
ratio, which is produced before the Universe becomes radiation
dominated;
\begin{equation}
\frac{n_B}{s}\sim B_0 \left( \frac{T_R}{T_B}\right)^5,
\end{equation}
where $T_B$ denotes the temperature when the baryon number asymmetry is 
produced.
In the most naive case, when $T_B\sim 10TeV$, and
$T_R\sim 1GeV$, the
dillution factor is crucially small.
If the potential for $\phi_{string}$ is sufficiently flat, which may
occur in supersymmetric extensions of the model, 
the time of the string formation is delayed and one can improve the result.
However, the supersymmetric extension seems far from being attractive
for scenarios with TeV scale unification.
Otherwise, one should fine tune the self-coupling constant,
which makes this scenario unsatisfactory.

One can consider another scenario for string formation.
Preheating after inflation\cite{PR_FKLT} may
lead to nonthermal phase transitions with defect formation.
The time of the string formation may be delayed so that the string formation
occurs just before the reheating, but it requires 
the fine tuning of the coupling constants, which makes this scenario
unattractive.}

\subsection{Baryogenesis from unstable Domain walls}
\hspace*{\parindent}
Unlike the strings that we have discussed above, unstable domain walls 
are the efficient source of baryon number asymmetry, if they decay after 
reheating. 

In most cases, the domain walls are dangerous for the standard evolution
of the Universe\cite{KZO}.
However, if some criteria are satisfied, unstable domain walls that
disappear at $t<t_c \sim (G\sigma)^{-1}$ can exist\cite{vilenkin}.
When the discrete symmetry is not the exact symmetry, it may be broken
by the interactions suppresses by the cut-off scale.
Then the degeneracy is broken and the energy difference 
$\epsilon\ne0$ appears\cite{matsuda_wall}.
Regions of the higher density vacuum tend to collapse when the pressure
induced by the energy difference becomes dominant.
The corresponding force per unit area of the wall is $\sim \epsilon$.
The energy difference $\epsilon$ becomes dynamically 
important when this force becomes comparable to the force of 
the tension $f\sim \sigma/R_{w}$,
where $\sigma$ is the surface energy density of the wall
and $R_{w}$ denotes the typical scale for the wall distance.
For walls to disappear, this has to happen before they become harmful.
On the other hand, the domain wall network is not a static 
system.
In general, the initial shape of the walls right after the
phase transition is determined by the random variation
of the scalar VEV.
One expects the walls to be very irregular, random
surfaces with a typical curvature radius, which
is determined by the correlation length of the
scalar field.
To characterize the system of domain walls,
one can use a simulation\cite{simulation}.
The system will be dominated by one large (infinite size)
wall network  and some finite closed walls (cells) when
they form.
The isolated closed walls smaller than the horizon
will shrink and disappear soon after the phase transition.
Since the walls smaller than the horizon size 
will efficiently disappear so that only walls
at the horizon size will remain, 
their typical curvature scale will be  the horizon 
size, $R\sim t\sim M_{p}/g_{*}^{\frac{1}{2}}T^{2}$.
Since the energy density of the wall $\rho_{w}$ is about
\begin{equation}
\rho_{w}\sim \frac{\sigma}{R},
\end{equation}
and the radiation energy density $\rho_{r}$ is 
\begin{equation}
\rho_{r}\sim g_{*}T^{4},
\end{equation}
one sees that the wall dominates the evolution 
below a temperature $T_{w}$
\begin{equation}
T_{w}\sim \left(\frac{\sigma}{g_{*}^{1/2}M_{p}}
\right)^{\frac{1}{2}}.
\end{equation}
To prevent the wall domination, one requires the
pressure to have become dominant before this epoch,
\begin{equation}
\label{criterion}
\epsilon>\frac{\sigma}{R_{wd}}\sim
\frac{\sigma^{2}}{M^{2}_{p}},
\end{equation}
which is consistent with the criterion in ref.\cite{KZO,vilenkin}.
Here $R_{wd}$ denotes the horizon size at the wall domination.
A pressure of this magnitude would be produced by
higher dimensional operators, which explicitly break
 the effective discrete symmetry\cite{matsuda_wall}.

The criterion (\ref{criterion})
seems appropriate, if the scale of the wall is higher 
than $(10^{5}GeV)^{3}$.
For the walls below this scale ($\sigma\le(10^{5}GeV)^{3}$),
 there should be  further constraints coming from primordial 
nucleosynthesis.
Since the time associated with the collapsing temperature
 $T_{w}$
is $t_{w}\sim M_{p}^{2}/g_{*}^{\frac{1}{2}}\sigma
\sim 10^{8}\left(\frac{(10^{2}GeV)^{3}}{\sigma}\right)$sec,
the walls $\sigma\le(10^{5}GeV)^{3}$ will decay after 
nucleosynthesis\cite{Abel}.
In this case, one must consider stronger constraint.

The cosmological domain wall that we consider in this paper will have
the surface energy density $\sigma \simeq M_*^3 \simeq (10^2 TeV)^3$,
which can decay after or just before the reheating.
When they decay, their energy density $\rho_w$ can be comparable to 
the energy density of the radiation.
Moreover, their baryon number violating decay is enhanced if the
center of the wavefunction of baryons and leptons along extra dimension
coincides in the core of the defect.
There may be other source of the enhancement in the false vacuum domain
where the center of the wavefunctions may be shifted by O(1).

We can calculate the baryon to entropy ratio as
\begin{equation}
\frac{n_B}{s}\simeq \frac{\rho_w /m_X}{T_d^3}\epsilon 
\le \frac{T_d}{m_X}\epsilon
\end{equation}
where $T_d$ denotes the temperature when walls decay.
The effective CP violation and the difference between the rates of the
decays is included in $\epsilon$.
The mass of $X$ boson that mediates the baryon number violating decay 
is assumed to be $m_X \simeq M_*$.
If the baryon number violating interaction is not enhanced in the
background of the wall configuration, the exponential suppression in
$\epsilon$ reduces the baryon number asymmetry, which is very similar to
the situation discussed in ref.\cite{thermalB}.
However, there should be a huge enhancement of the baryon number violating
interactions in our model.
In the most optimistic case, where $\epsilon \simeq 10^{-4}$,
one can obtain the desired baryon to entropy ratio for 
$T_d \simeq 10^{-1}GeV$ and $m_X \simeq 10^5 GeV$.

\section{Conclusions and Discussions}
\hspace*{\parindent}
In the standard grand unified theories, various types of
baryon number violating processes are expected in the early Universe.
On the other hand, in models with large extra dimensions, it is
difficult to realize baryon number violation even in the background of a
string. 
In this paper we argue that in the background of a cosmic string with
extra dimensions the baryon number violating interactions are enhanced
if the proton decay is suppressed by the machanism of 
localized wavefunctions. 
We also make some comments on baryogenesis mediated by cosmological
defects.  
At least two scenarios will be successful.
One is the scenario of leptogenesis where the required lepton number
conversion is mediated by the string, and the other is the baryogenesis
from the decaying cosmological domain wall. 
Both scenarios are new and are not discussed in the past.
These mechanisms predicts sufficient baryon number production even if
the reheating temperature is much lower than the temperature of the
electroweak phase transition.

\section{Acknowledgment}
\hspace*{\parindent}
We wish to thank K.Shima for encouragement, and our colleagues in
Tokyo University for their kind hospitality.

\end{document}